\documentclass{jpp}

\usepackage{graphicx}
\usepackage{caption}
\usepackage{subcaption}
\usepackage{dcolumn}
\usepackage{bm}
\usepackage[dvipsnames]{xcolor}
\usepackage{amssymb}
\usepackage{amsmath}
\usepackage{verbatim}
\usepackage{appendix}
\usepackage{listings}
\usepackage{units}
\usepackage{csquotes}
\usepackage{tikz}

\usepackage{hyperref}

\usepackage{etoolbox}
\makeatletter
\patchcmd{\hyper@makecurrent}{%
    \ifx\Hy@param\Hy@chapterstring
        \let\Hy@param\Hy@chapapp
    \fi
}{%
    \iftoggle{inappendix}{
        \@checkappendixparam{chapter}%
        \@checkappendixparam{section}%
        \@checkappendixparam{subsection}%
        \@checkappendixparam{subsubsection}%
        \@checkappendixparam{paragraph}%
        \@checkappendixparam{subparagraph}%
    }{}%
}{}{\errmessage{failed to patch}}

\newcommand*{\@checkappendixparam}[1]{%
    \def\@checkappendixparamtmp{#1}%
    \ifx\Hy@param\@checkappendixparamtmp
        \let\Hy@param\Hy@appendixstring
    \fi
}
\makeatletter

\newtoggle{inappendix}
\togglefalse{inappendix}

\apptocmd{\appendix}{\toggletrue{inappendix}}{}{\errmessage{failed to patch}}
\apptocmd{\subappendices}{\toggletrue{inappendix}}{}{\errmessage{failed to patch}}

\newcommand*\obar[2][0.75]{
    \sbox{\myboxA}{$\m@th#2$}%
    \setbox\myboxB\null
    \ht\myboxB=\ht\myboxA%
    \dp\myboxB=\dp\myboxA%
    \wd\myboxB=#1\wd\myboxA
    \sbox\myboxB{$\m@th\overline{\copy\myboxB}$}
    \setlength\mylenA{\the\wd\myboxA}
    \addtolength\mylenA{-\the\wd\myboxB}%
    \ifdim\wd\myboxB<\wd\myboxA%
       \rlap{\hskip 0.5\mylenA\usebox\myboxB}{\usebox\myboxA}%
    \else
        \hskip -0.5\mylenA\rlap{\usebox\myboxA}{\hskip 0.5\mylenA\usebox\myboxB}%
    \fi}
\makeatother

\definecolor{colorexample}{RGB}{250,143,56}

\definecolor{ecolor}{HTML}{4477AA}
\definecolor{Hcolor}{HTML}{117733}
\definecolor{Necolor}{HTML}{88CCEE}
\definecolor{Hecolor}{HTML}{DDCC77}
\definecolor{Ccolor}{HTML}{CC6677}

\newcommand{\ordo}[1]{{\cal O}\left( #1 \right)}

\renewcommand{\vec}[1]{\boldsymbol{#1}}
\newcommand{\im}{\ensuremath{\mathrm{i}}}
\newcommand{\e}{\ensuremath{\mathrm{e}}}

\renewcommand{\d}{\ensuremath{\mathrm{d}}}

\newcommand{\remove}[1]{{}}

\newcommand*{\nouncite}[1]{\citet{#1}}
\newcommand{\lang}{\left\langle}
  \newcommand{\rang}{\right\rangle}

\newcommand{\appref}[1]{\hyperref[#1]{Appendix~\ref{#1}}}

\usepackage{morefloats}

\title{Effects of collisions on impurity transport driven by electrostatic modes}
\author{S.~Buller\aff{1}\corresp{\email{stefan.buller@ipp.mpg.de}} , P.~Helander\aff{1}} 
\affiliation{\aff{1}Max Planck Institute for Plasma Physics, Greifswald, Germany}

\begin{document}

\maketitle

\begin{abstract}
  The turbulence-induced quasi-linear particle flux of a highly-charged, collisional impurity species is calculated from the electrostatic gyrokinetic equation including collisions with the bulk ions and the impurities themselves. The equation is solved by an expansion in powers of the impurity charge number $Z$.
  In this formalism, the collision operator only affects the impurity flux through the dynamics of the impurities in the direction parallel to the magnetic field. At reactor-relevant collisionality, the parallel dynamics is dominated by the parallel electric field, and collisions have a minor effect on the turbulent particle flux of highly-charged, collisional impurities. 

\end{abstract}


\section{Introduction}
Impurities are always present in fusion plasmas, either due to unavoidable plasma-wall interaction, or through deliberate impurity injection. In the edge of a tokamak or stellarator, impurities can be beneficial, as they radiate energy and thus can mitigate the heat-load on plasma-facing components. However, their ability to radiate energy is detrimental in the core of the plasma. It is thus crucial to understand how impurities are transported so that they do not accumulate in the core of the device.

There is a large body of theoretical research on the neoclssical transport of impurities in stellarators \citep{velasco2017, helanderPRL2017, calvo2018nf}, which we shall not describe in detail. Far less has been done to study turbulent particle transport - either of impurities or of the bulk ions and electrons -- and most of these studies rely on quasi-linear transport theory \citep{mikkelsen2014}. Recently there have however been direct numerical simulations of turbulence in impure stellarator plasmas using gyrokinetic codes \citep{nunami2020}.

Recent measurement in the stellarator Wendelstein 7-X indicate that the impurity transport is dominated by turbulent diffusion \citep{langenberg2018,bgeiger2019}. 
Diffusion coefficients two orders of magnitude larger than those calculated from collisional transport have been measured for iron impurities \citep{bgeiger2019}, and the impurity confinement time appears to be insensitive to the impurity charge number \citep{langenberg2020} -- in contradiction to predictions for collisional transport \citep{helander2005}.

On the other hand, the experimental observations may be consistent with recent theoretical calculations of the transport due to electrostatic turbulence \citep{helander2018}, which, for heavy species, give transport coefficients independent of the impurity charge and mass \citep{helander2018,angioni2016}. However, the calculation by \nouncite{helander2018} does not include collisions, which could have a significant effect on heavy impurities due to their high charge and high collision frequency. The present paper addresses this shortcoming by including collisions in the calculation of the quasi-linear impurity flux. 

Previous analytical work has shown that collisions with the impurities themselves do not significantly affect the impurity flux in tokamaks \citep{pusztai13}. This calculation can be generalized, without additional complications, to also apply to stellarators, which is done in \autoref{eq:Czzonly}.
However, \nouncite{pusztai13} considered non-trace impurities, which allowed them to neglect the collisions between impurities and bulk ions, which could, in principle, modify the impurity flux.
In \autoref{Call}, we show that impurity-ion collisions provide only a small correction to the previous results unless the charge number of the impurities is comparable to the inverse bulk-ion collisionality. Our result thus strengthens the conclusions of \nouncite{helander2018} and \nouncite{pusztai13} for low collisionality plasmas, and generalizes parts of the calculation of \nouncite{pusztai13} to stellarator geometry.


\section{Equation for heavy impurities\label{sec:eq}}
The linearized electrostatic gyrokinetic equation for impurities is
\begin{equation}
\im v_\| \nabla_\| \hat{g}_z(l,v,\lambda) + (\omega - \omega_{\text{d}z}) \hat{g}_z(l,v,\lambda) - \im C[\hat{g}_z(l,v,\lambda)] =  (\omega - \omega_{*z}^T) \frac{ZeJ_0\hat{\phi}}{T_z} f_{Mz}, \label{eq:gke}
\end{equation}
where $g_z$ is the non-adiabatic part of the perturbed impurity distribution function, $g_z = f_z - (1 - Ze\phi/T_z)f_{Mz}$, $f_z$ the full impurity distribution function, $f_{Mz}$ a Maxwellian with temperature $T_z$ and density $n_z$,
\begin{equation}
f_{Mz} (v) = n_z \left(\frac{m_z}{2\pi T_z}\right)^{3/2} \e^{-\frac{m_z v^2}{2T_z}},
\end{equation}
with $m_z$ the mass of the impurity.
$\phi$ the fluctuating electrostatic potential, and $e$ the elementary charge. Both $g_z$ and $\phi$ have been written as $g_z = \hat{g}_z(l)\e^{-\im \omega t +\im S}$, where (in ballooning space) $S$ satisfies $\vec{B} \cdot \nabla S = 0$ and $\nabla S = \vec{k}_\perp$, where $\vec{B}$ is the magnetic field and $l$ is the arc length along $\vec{B}$. The magnetic field is written as $\vec{B} = \nabla \psi \times \nabla \alpha$, and the wave-vector as $\vec{k}_\perp = k_\psi \nabla \psi + k_\alpha \nabla \alpha$.
The drift-frequency is $\omega_d = \vec{k}_\perp \cdot \vec{v}_d$, where $\vec{v}_d$ is the drift-velocity
\begin{equation}
 \vec{v}_{\text{d}z} = \frac{v_\perp^2}{2\Omega_z} \vec{b} \times \nabla \ln{B} + \frac{v_\|^2}{\Omega_z} \vec{b} \times \left(\vec{b} \cdot \nabla \vec{b}\right),
\end{equation}
where $\vec{b} = \vec{B}/B$, $B = |\vec{B}|$, $\Omega_z = ZeB/m_z$; $v_\|$ and $v_\perp$ the speed in the direction parallel and perpendicular to $\vec{B}$. 
The collision operator $C$ will be specified explicitly in \autoref{sec:g12}. The diamagnetic frequency is $\omega_{*z}^T = \omega_{*z} (1 + \eta_z [x^3 - 3/2])$, where $
\omega_{*z} = (k_\alpha T_z/Z e)\, \d \ln n_z/\d \psi$, $\eta_z = \d \ln T_z/ \d \ln n_z$ and $x = v/v_{Tz}$ with $v_{Tz} = \sqrt{2T_z/m_z}$. $J_0 = J_0(k_\perp v_\perp/\Omega_z)$, where $J_0$ is the zeroth order Bessel function of the first kind. In \eqref{eq:gke} and throughout the rest of this paper, gradients are taken with $\lambda=v_\perp^2/(Bv^2)$ and $v$ fixed.
The derivation of \eqref{eq:gke} assumes that the electrostatic potential perturbations have low amplitude, in the sense that $Ze \hat{\phi}/T_z \ll 1$, and that the deviation of the background potential from being a flux-function is similarly small. If these conditions are violated, the potential variations would cause $n_z$ to vary on the flux-surface, and \eqref{eq:gke} would not be valid. Such large variations in the background potential have been observed experimentally and studied theoretically for both tokamaks \citep{reinke2012, fulop2011} and stellarators \citep{pedrosa2015, garcia2017}, but will not be considered here.

Given a solution to \eqref{eq:gke}, we calculate the quasi-linear impurity flux using \citep{helander2018}
\begin{equation}
\Gamma_z = - k_\alpha \mathcal{I} \lang \int\! \d^3 v\, \hat{\phi}^* J_0 \hat{g}\rang, \label{eq:qlf}
\end{equation}
where $\mathcal{I}$ denotes the imaginary part and the brackets denote a flux-surface average
\begin{equation}
\lang X \rang = \lim_{L \to \infty} \left.\int_{-L}^L \!X \frac{\d l}{B} \middle/ \int_{-L}^L \!\frac{\d l}{B}\right..
\end{equation}

\subsection{Expansion in powers of $Z^{-1}$}
Like \nouncite{pusztai13}, we solve \eqref{eq:gke} for a highly-charged impurity species by expanding the equation in powers of $Z$. We assume
\begin{align}
  Z^2 \frac{n_z}{n_e} &\ll 1, \\
  Z^{1/2} &\gg 1, \\
 \frac{m_z}{m_i} \sim Z &\gg 1, 
\end{align}
corresponding to a highly-charged, heavy trace impurity species. As the impurities are only a trace, they will not affect the electrostatic potential, which is set by the bulk ions and electrons. Thus, we assume that the impurities merely respond to ion-scale turbulence, and that $\omega$ is comparable to the ion diamagnetic frequency $\omega_{*i}$. We order the impurity frequencies in powers of $Z$ by relating them to the corresponding bulk ion frequencies
\begin{align}
  &\omega_{bz} \sim Z^{-1/2} \omega_{bi}, \\
  &\omega_{dz} \sim Z^{-1}\omega_{di},
\end{align}
where we order the bulk ion frequencies as similar $\omega_{bi} \sim \omega_{di}$. 
Here, $\omega_{ba}$ is the bounce or transit frequency of species $a$, $v_\| \nabla_\| g_a \sim \omega_{ba} g_a$.
The collision operator is ordered as
\begin{align}
  &C[g_z] \sim \omega_{*i} g_z.
\end{align}
As the turbulence is set by the bulk species, $k_\perp$ is independent of $Z$ and $k_\perp v_\perp \Omega_z$ thus scales as $Z^{-1}$.

We expand $\hat{g}_z$ and \eqref{eq:gke} in powers of $Z^{-1}$,
\begin{equation}
\hat{g}_z = \hat{g}_z^{(0)} + \hat{g}_z^{(1/2)} + \hat{g}_z^{(1)} + \dots,
\end{equation}
where $\hat{g}_z^{(n)}/\hat{g}_z^{(0)} \sim Z^{-n}$.

The $Z^0$ order equation becomes
\begin{equation}
  \omega \hat{g}^{(0)}_z - \im C[\hat{g}_z^{(0)}] =  \omega \frac{Ze\hat{\phi}}{T_z} f_{Mz}, \label{eq:gke0}
\end{equation}
which has the solution $\hat{g}_z^{(0)} = (Ze\hat{\phi}/T_z)\, f_{Mz}$, and gives no impurity flux when inserted into \eqref{eq:qlf}.

The $Z^{-1/2}$ order equation is
\begin{equation}
\omega  \hat{g}_z^{(1/2)} - \im C[\hat{g}_z^{(1/2)}] = -\im v_\| \nabla_\| \hat{g}_z^{(0)}, \label{eq:gke12}
\end{equation}
so that also $\hat{g}_z^{(1/2)}$ yields no flux when inserted into \eqref{eq:qlf}, since
\begin{equation}
  \int \!\d^3 v\, C[g_z] = 0,
\end{equation}
for collisions that preserve the number of impurities.

The $Z^{-1}$ order equation is
  \begin{align}
    &\omega  \hat{g}_z^{(1)} - \im C[\hat{g}_z^{(1)}]    \label{eq:gke1}
\\
    =&   \omega_{\text{d}z} \hat{g}_z^{(0)} - \im v_\| \nabla_\| \hat{g}_z^{(1/2)} - \omega_{*z}^T \frac{Ze\hat{\phi}}{T_z} f_{Mz} - \omega \frac{k_\perp^2 v_\perp^2}{4\Omega_z^2}\frac{Ze\hat{\phi}}{T_z} f_{Mz}, \nonumber
  \end{align}
where we have expanded the Bessel function $J_0$. The corresponding particle flux receives contributions from all but the last term on the right-hand side
  \begin{equation}
    \Gamma_z = -k_\alpha  
    \mathcal{I} \lang \int \!\!\d^3 v\, \hat{\phi}^*  \left(\frac{\omega_{\text{d}z}}{\omega} \hat{g}_z^{(0)} - \frac{\im v_\|}{\omega} \nabla_\| \hat{g}_z^{(1/2)}
    \vphantom{\frac{Ze\hat{\phi}}{T_z}}
    \right.\right.
- \frac{\omega_{*z}^T}{\omega} \frac{Ze\hat{\phi}}{T_z} f_{Mz} \left.\left. \vphantom{\frac{Ze\hat{\phi}}{T_z}}\right)\rang.  \label{eq:qlf1}
  \end{equation}
  The terms in this expression are fluxes due to magnetic curvature, parallel compressibility \citep{angioni2006}, and ordinary diffusion. The curvature and diffusion terms were also found by \nouncite{helander2018}, along with a thermodiffusion term, which is smaller in $Z^{-1}$ and thus absent to this order. The parallel compressibility term is absent in the calculation of \nouncite{helander2018}, since no distinction was made between the smallness of the impurity bounce- and drift-frequency, but our results otherwise agree with those of \nouncite{helander2018}.
  
  In \eqref{eq:qlf1}, collisions give no direct contribution to the flux, but will affect the flux indirectly through $g_z^{(1/2)}$ in the parallel compressibility term. 
  To quantify the effects of collisions, we thus have to solve \eqref{eq:gke12} for $g_z^{(1/2)}$, which we do in the next section. Once an expression for $g_z^{(1/2)}$ has been obtained, we can then estimate the importance of collisions by comparing the flux due to the parallel compressibility with the ordinary diffusive flux.

\section{Solving for $\hat{g}_z^{(1/2)}$\label{sec:g12}}
Before solving \eqref{eq:gke12}, we note that only the part of $\hat{g}_z^{(1/2)}$ that is odd in $v_\|$ will contribute to the flux \eqref{eq:qlf1}. We thus split \eqref{eq:gke12} into an odd and even part, where the odd part is
\begin{equation}
\omega  \hat{g}_z^{(1/2)-} - \im C^{-}[\hat{g}_z^{(1/2)}] = -\im v_\| \nabla_\| \hat{g}_z^{(0)}, \label{eq:gke12m}
\end{equation}
where the ``$-$'' superscript indicates the part of $\hat{g}_z$ and $C[\hat{g}_z^{(1/2)}]$ that is odd in $v_\|$. 

To solve \eqref{eq:gke12m}, we need an explicit expression for the collision operator. We write
\begin{equation}
C[\hat{g}_z] = C_{zz}[\hat{g}_z] + C_{zi}[\hat{g}_z], \label{eq:CzzCzi}
\end{equation}
where $C_{zz}$ and $C_{zi}$ are the impurity-impurity and impurity-ion collision operators, respectively. In the limit where finite Larmor-radius effects can be neglected, we can use the expressions for the Fokker-Planck collision operator from collisional transport theory directly on $\hat{g}_z$. This is easily justifiable for the impurities, which have a small Larmor radius due to their large charge. For the reminder of this paper, we thus simplify the notation by omitting the hats on $g_z$ and $\phi$. 

The relative size of the impurity-impurity and impurity-ion operators is \citep{helander2005}
\begin{equation}
\frac{C_{zz}[g_z]}{C_{zi}[g_z]} \sim \sqrt{\frac{m_z}{m_i}} Z^2 \frac{n_z}{n_e}, \label{eq:CzzCziratio}
\end{equation}
and will be taken to be $\ordo{1}$ in our orderings.
For purely illustrative purposes, it is nevertheless instructive to consider the limit where $C_{zz}[g_z]\gg C_{zi}[g_z]$, to demonstrate why impurity-impurity collisions cannot affect the impurity flux. 

\subsection{Impurity-impurity collisions only\label{eq:Czzonly}}

For $C[g_z^{(1/2)}] \approx C_{zz}[g_z^{(1/2)}]$, $g_z^{(1/2)} \propto v_\| f_M$ is in the null-space of the collision operator \citep{helander2005}, in the sense that $C_{zz}[v_\| f_M] = 0$. The solution to \eqref{eq:gke12m} then becomes
\begin{equation}
g_{z}^{(1/2)-} = -\frac{\im}{\omega} v_\| \frac{Ze f_{Mz}}{T_z}\nabla_\| \phi. \label{eq:g12mm}
\end{equation}
This result, previously found by \nouncite{pusztai13}, would also have been obtained from \eqref{eq:gke12m} without the collision operator, and is thus not affected by impurity-impurity collisions. 

Inserting \eqref{eq:g12mm} into \eqref{eq:qlf1}  and writing out $\omega = \omega_r + \im \gamma$, the parallel compressibility contribution to the particle flux becomes
\begin{equation}
  \begin{aligned}
    \Gamma_z^{\text{comp}}= & k_\alpha \mathcal{I}\frac{Ze}{T_z}\lang \int f_{Mz} \frac{v_\|}{\omega^2}  \nabla_\|\left(  v_\| \nabla_\| \phi\right) \phi^* \d^3 v \rang \\
    = &-k_\alpha\frac{Ze}{m_z} n_z \lang |\nabla_\| \phi|^2 \rang  \mathcal{I} \frac{1}{\omega^2}. \\
    = &\frac{2\omega_r \gamma k_\alpha }{(\omega_r^2 + \gamma^2)^2}\frac{Ze}{m_z} n_z \lang |\nabla_\| \phi|^2 \rang.
  \end{aligned}
  \label{eq:relevantflux}
\end{equation}
 The right-hand side of \eqref{eq:relevantflux} follows from $\langle B\nabla_\| X \rangle = 0$ for any single-valued $X$; also recall that $\lambda$ and $v$ are kept fixed when evaluating $\nabla v_\|$. We compare this flux to the flux due to ordinary diffusion. From the last term in \eqref{eq:qlf1}, the diffusive flux is of the size
\begin{equation}
\Gamma_z^{\text{D}} \sim - \frac{k_\alpha^2}{\omega} \lang |\phi|^2\rang  \frac{\d n_z}{\d \psi},
\end{equation}
whereupon the relative contribution of \eqref{eq:relevantflux} to the flux becomes
\begin{equation}
\hspace*{-1ex}\frac{\Gamma_z^{\text{comp}}}{\Gamma_z^{\text{D}}}  \sim \frac{Ze}{m_z} \frac{k_\|^2}{k_\alpha \omega \frac{\d \ln n_z}{\d \psi}} \sim \left(\frac{k_\| a}{k_\perp \rho_i}\right)^2, \label{eq:ratio}
\end{equation}
where we have used $\nabla_\| \phi \sim k_\| \phi$, recalled $\omega \sim \omega_{*i} \sim k_\perp \rho_i v_{Ti}/a$, used $k_\alpha \sim k_\perp a$ and $\d \ln n_z/\d \psi \sim 1/(Ba^2)$. Here, $a$ is the minor radius.
For a stellarator with $N$ field-periods and major radius $R$, we can use the rough estimate
\begin{equation}
  k_\| \sim \frac{N}{2R}, \label{eq:kpar}
\end{equation}
based on the picture of $\phi$ as a standing wave on each period of the stellarator \citep{helander2012,kornilov2004}.
The ratio \eqref{eq:ratio} thus scales as the inverse aspect-ratio squared.
For parameters typical of ion-temperature-gradient turbulence in Wendelstein 7-X ($k_\perp \rho_i \sim 10^{-1}$ to $10^0$; $k_\| a \sim 2.5 \times 10^{-1}$), the ratio \eqref{eq:ratio} is about  $5$ to $5 \times 10^{-2}$, where the larger value corresponds to turbulence with smaller $k_\perp \rho_i$. 

We thus conclude that it could make a significant contribution to the impurity flux.
In the next section, we show how this contribution is modified by impurity-ion collisions. 

\subsection{Effects of impurity-ion collisions\label{Call}}
We have shown that impurity self-collisions have no effect on the quasi-linear particle flux of highly-charged impurities. This result is applicable in the limit where highly-charged impurities are a trace $Z^2 n_z/n_i \ll 1$ with exceptionally large mass, $\sqrt{m_z/m_i} \gg n_i/(Z^2n_z)$, and hinges on the fact that the solution to \eqref{eq:gke12m} without impurity-impurity collisions is in the null-space of $C_{zz}$. To generalize these results to impurities without exceptionally large mass, we need to include the effects of impurity-ion collisions. 

Neglecting finite Larmor-radius effects, the impurity-ion collision operator is (to lowest order in $\sqrt{m_i/m_z}$) \citep{calvo2019} 
\begin{equation}
  \begin{aligned}
    & C_{zi}^-[g_z^{(1/2)}] \\
    =& \frac{4}{3\sqrt{\pi}}\sqrt{\frac{m_i}{m_z}}\hat{\nu}_{zi}  \left(\mathcal{K}[g_z^{(1/2)-}] + \frac{m_z v_\| A}{T_z} f_{Mz}\right);
  \end{aligned} \label{eq:Coperator}
\end{equation}
where
\begin{equation}
 \mathcal{K}[g] = \frac{T_z}{m_z} \nabla_v \cdot \left[f_{Mz} \nabla_v \left(\frac{g}{f_{Mz}}\right)\right],
\end{equation}
with $\nabla_v$ the gradient operator in velocity space; and
\begin{equation}
A  = \frac{3\sqrt{\pi} T_z^{3/2}}{\sqrt{2} n_i m_i^{3/2}} \int \frac{v_\|}{v^3} f_i(\vec{v}) \d^3 v,
\end{equation}
where $f_i$ is the bulk ion distribution. $A$ can be interpreted as the flow velocity the impurities would reach due to collisions with the bulk ions, in absence of other forces \citep{calvo2019}. The collision frequency is
\begin{equation}
  \hat{\nu}_{ab} = \frac{Z_a^2 Z_b^2 n_b}{m_a^{1/2}T_a^{3/2}} \frac{e^4 \ln\Lambda}{2^{3/2}4\pi  \epsilon_0^2},
\end{equation}
with $\ln \Lambda$ the Coulomb-logarithm and $\epsilon_0$ the permittivity of vacuum. To simplify the notation, we also introduce the modified impurity-ion collision frequency
\begin{equation}
\nu_{zi}' = \frac{4}{3\sqrt{\pi}}\sqrt{\frac{m_i}{m_z}}\hat{\nu}_{zi}.
\end{equation}
The operator \eqref{eq:Coperator} is a mass-ratio expanded Fokker-Planck operator; the general Fokker-Planck operator implemented in several gyrokinetic codes \citep{cgyro2016,pan2019} should thus reduce to the above operator in the appropriate limit.


With $C=C_{zz} + C_{zi}$ in \eqref{eq:gke12m}, the Ansatz $g^{(1/2)-}_z \propto v_\| f_{Mz}(v)$ yields 
\begin{equation}
g^{(1/2)-}_z = -\frac{\im v_\|}{T_z} \frac{f_{Mz}}{ \omega  + \im \nu_{zi}'}\left(Ze \nabla_\| \phi - \nu_{zi}' m_z A\right), \label{eq:g12mmm}
\end{equation}
where we have used $C_{zz}[v_\| f_{Mz}] =0$ and $\mathcal{K}[v_\| f_{Mz}] = -v_\| f_{Mz}$.
Note that impurity-impurity collisions again have no effect, as the solution is in the null-space of $C_{zz}$. Impurity-ion collisions, on the other hand, both modify the response to the parallel electric field, and provide a new source for $g^{(1/2)-}_z$ through the friction force between the impurities and bulk ions.

The relative size of the ion-impurity friction and the electric field terms in \eqref{eq:g12mmm} is, assuming $e \phi/T_z \sim \rho_i/a$, $A \sim \rho_i v_{Ti}/a$ (appropriate since $A$ is a flow velocity),
\begin{equation}
 \frac{\nu_{zi}' m_z  A }{ Ze \nabla_\| \phi}  \sim Z \frac{a \hat{\nu}_{ii} }{v_{Ti}} \frac{1}{(k_\| a)}\label{eq:par},
\end{equation}
which is essentially $Z$ times the bulk-ion collisionality divided by $k_\| a$. As any fusion reactor will be in a low collisionality regime $a \hat{\nu}_{ii}/v_{Ti} \ll 1$, the above ratio will likely be small.
However, it can be significant in smaller fusion experiments, such as TJ-II, as shown in \autoref{tab:values}. 
Likewise, the effect of the $\im \nu_{zi}'$ in the denominator can be estimated as
\begin{equation}
\frac{\nu_{zi}'}{\omega} \sim Z \frac{\hat{\nu}_{ii} a}{v_{Ti}} \frac{1}{(k_\perp \rho_i)}, \label{eq:colr2}
\end{equation}
which again scales as $Z$ times the bulk-ion collisionality.
Thus, in the limit where $\Gamma_z^{\text{comp}}$ is significant (the ratio \eqref{eq:ratio} is large, $k_\| a > k_\perp \rho_i$), the collisional modification of the response to $\nabla_\| \phi$ in \eqref{eq:g12mmm} is more important than the drive due to ion-impurity friction, but both of these modifications are likely small in the Large Helical Device and Wendelstein 7-X, and will be yet smaller in a fusion reactor.



\begin{table}
  \begin{center}
    \def~{\hphantom{0}}
    \begin{tabular}{lllll}
      Scenario & $T_i/\mathrm{keV}$ & $n_i/\mathrm{10^{-20}m^{-3}}$ & $\hat{\nu}_{ii} a/v_{Ti}$ & $Z_{\text{u}}$\\[3pt]
      W7-X &  1.1 & 0.3 & $1.5 \times 10^{-3}$ & 200\\ 
      TJ-II &  0.1 & 0.07 & $17 \times 10^{-3}$ & 18\\
      LHD &  1.5 & 0.4 & $1.2 \times 10^{-3}$ & 250\\
    \end{tabular}
    \caption{\label{tab:values} Collisionality $\hat{\nu}_{ii} a/v_{Ti}$ calculated for different scenarios. $Z_{\text{u}} \equiv 0.3v_{Ti}/(\hat{\nu}_{ii} a)$ refers to the charge number at which impurity-ion collisions are expected to have an order unity effect on the flux due to parallel compressibility \eqref{eq:relevantflux4}. The parameters are taken form the following scenarios: W7-X -- LBO impurity study \citep{langenberg2020}; TJ-II -- LBO impurity study \citep{zurro2014}; LHD -- TESPEL impurity study \citep{tamura2016}. In $Z_u$ we used $k_\| a \sim k_\perp \rho_i \sim 0.3$ to obtain one estimate for both \eqref{eq:par} and \eqref{eq:colr2}.}
  \end{center}
\end{table}

Including both of the impurity-ion collisional modifications, the parallel compressibility flux becomes
\begin{equation}
    \hspace*{-3ex}\Gamma_z^{\text{comp}}= -k_\alpha \mathcal{I} \frac{1}{\omega (\omega + \im \nu_{zi}')}\frac{Ze n_z }{m} \lang   |\nabla_\| \phi|^2 + A\nabla_\| \phi^*\rang.
  \label{eq:relevantflux4}
\end{equation}
It is difficult to draw any detailed conclusions from this expression, as the $A\nabla_\| \phi^*$ term causes the flux to both depend on the phase of the imaginary $\phi$ and the ion-impurity friction force, which are beyond the scope of this work. 

\section{Summary \& conclusions}
We have included impurity-ion collisions in the calculation of the quasi-linear particle flux of highly-charged impurities. The lack of collisions was thought to be one of the main shortcomings of previous analytical calculations \citep{helander2018}, and it can indeed affect the impurity flux if the bulk ion collisionality times the charge number of the impurity is not small. This effect could thus be significant in present days experiments, in particular experiments with low ion temperature, such as TJ-II, but is not expected to be important in a fusion reactor or in larger fusion experiments.

As this result was based on an expansion in the largeness of the impurity charge number, it is not applicable to species with low charge -- such as carbon -- at least not in a quantitative sense. Indeed, for electrons, collisions can have a large effect on the electron particle transport\citep{angioni2005, fulop2008}, especially at low $k_y \rho_i$-values \citep{angioni2009}, if the electron-ion collision frequency is comparable to the mode frequency and/or drift frequency.
However, for a highly charged species, the distribution function is predominantly set by the local value of the electrostatic potential and its parallel derivative, and collisions have a small effect. 

There are a few extensions to this work that may modify the above conclusion:

Firstly, if the impurities are not a trace, their distribution would affect the electrostatic potential fluctuation through the quasi-neutrality equation, and the potential would have to be expanded in $Z^{-1}$. However, the effect of collisions would then be smaller, as impurity self-collisions would dominate over ion-impurity collisions, according to \eqref{eq:CzzCziratio}. Thus, the conclusions of this paper apply even more strongly to highly-charged non-trace impurities, as noted in \nouncite{pusztai13}. Of course, impurities would also affect the turbulence itself, but such effects are beyond the scope of the present paper.

Secondly, if the background impurity density were to vary on the flux-surface, the Maxwellian in \eqref{eq:g12mm} and \eqref{eq:g12mmm} would weight different parts of the flux-surface differently, which could affect the relative importance of the impurity-ion friction and the parallel electric field.

Lastly, collisions also play an important role in saturating nonlinear gyrokinetic turbulence\citep{krommes1999,schekochihin2008}, which has not been considered in this work.

\bibliographystyle{jpp}
\bibliography{plasma-bib} 

\end{document}